# Performance Comparison of Intrusion Detection Systems and Application of Machine Learning to Snort System

**Syed Ali Raza Shah and Biju Issac**
*School of Computing, Media and the Arts, Teesside University, England, UK*

## ABSTRACT

This study investigates the performance of two open source intrusion detection systems (IDSs) namely Snort and Suricata for accurately detecting the malicious traffic on computer networks. Snort and Suricata were installed on two different but identical computers and the performance was evaluated at 10 Gbps network speed. It was noted that Suricata could process a higher speed of network traffic than Snort with lower packet drop rate but it consumed higher computational resources. Snort had higher detection accuracy and was thus selected for further experiments. It was observed that Snort triggered a high rate of false positive alarms. To solve this problem a Snort adaptive plug-in was developed. To select the best performing algorithm for the Snort adaptive plug-in, an empirical study was carried out with different learning algorithms and Support Vector Machine (SVM) was selected. A hybrid version of SVM and Fuzzy logic produced a better detection accuracy. But the best result was achieved using an optimized SVM with the firefly algorithm with FPR (false positive rate) as 8.6% and FNR (false negative rate) as 2.2%, which is a good result. The novelty of this work is the performance comparison of two IDSs at 10 Gbps and the application of hybrid and optimized machine learning algorithms to Snort.

## 1. INTRODUCTION

Today many businesses rely on computer networks. These networks fulfil the needs of business, enterprises and government agencies to build knowledgeable, complicated information networks which integrate various technologies such as distributed data storage systems, encryption techniques, voice over IP (VoIP), remote or wireless access and web services. These computer networks have become more important as business partners access the information through extranets, customers communicate using networks through e-commerce transactions or Customer Relation Management (CRM) systems and employees connect with enterprise networks through virtual private networks (VPNs). These well-travelled paths make computer networks more vulnerable than ever before because today's attackers are well organized as they have time, expertise and resources to launch the attacks that can avoid detection by even secure networks.

The attackers act like normal users, generate data and hide their malicious activities under terabytes of data. They know that many security mechanisms cannot protect the networks due to the large amount of data stored, scalability issues or due to the lack of detection capabilities. The enterprises and government agencies need to monitor their network traffic to detect malicious activities and perform analysis to differentiate the malicious and legitimate user activities to protect their networks. Detecting malicious activities require intrusion detection systems (IDS) and in today's secure ICT infrastructure, the IDSs are part of most networks. However, the IDSs are only good if they have elite detection capabilities. It is critical that an IDS detection mechanism is accurate enough to differentiate between legitimate and malicious traffic that enter and leave the network. The possible results of using an IDS are as follows: detected malicious traffic (real alarms), undetected malicious traffic, legitimate traffic that IDS detect as malicious (false alarms) and legitimate traffic that IDS detect as good.

The elite IDSs detect as much malicious traffic as possible and reduce the number of false alarms. There are a number of commercial IDSs available in the market such as Juniper, McAfee, Cisco, Symantec etc. [1]. The commercial IDS generally do not provide the ideal performance as advertised and could compromise computer network security. Like the commercial IDSs, there are a number of open source IDSs available such as Snort, Suricata and Bro.

Snort and Suricata [2] were chosen for our study as we felt they have comparable functions, detection rule sets and syntax. They are both under GNU GPL license. They both support intrusion prevention system (IPS) feature and support medium to high-speed network, though Suricata is more scalable with its multi-threaded architecture. Both support IPv6 traffic and their installation and deployment are easy. In contrast, Bro is a flexible script based IDS and its policy scripts or rules are written in its own Bro scripting language that does not rely on traditional signature detection [2]. It is under BSD license and does not support IPv6 traffic. Installation of Bro can be difficult. Unlike Snort or Suricata, Bro does not offer inline intrusion prevention features. Both Snort and Suricata have similar features such as a module to capture the network packets, a module to decode and classify the network packets and a module to detect accurately the malicious or legitimate packets based on a rule set defined by both IDSs. Snort and Suricata inspect network packets for possible malicious traffic through the rule set and trigger alarms when the packet payload matches with one of the rules [3].

The Snort IDS has been in development since 1998 by Sourcefire and has become the de-facto standard for IDSs over the last decade and has been extensively deployed and investigated in research studies. Snort has a single threaded architecture as shown in Figure 1 which uses the TCP/IP stack to capture and inspect network packets payload. [4]. Snort has added a multi-instance feature to its 2.9 release to address the limitation of single-thread and has hinted that version 3.0 will be multithreaded by default.

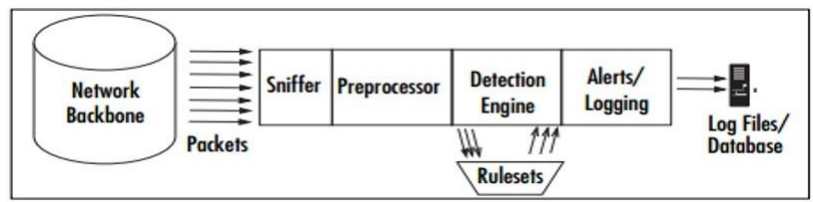

**Figure 1: Snort single threaded architecture**

The Suricata IDS was developed in 2010 by the Open Information Security Foundation (OISF). Suricata is publicised as a future next generation IDS integrating new ideas such as multithreading as shown in Figure 2. Based on the previous research it has improved on Snort because it uses multi-thread architecture to quickly capture and decode network packets [5].

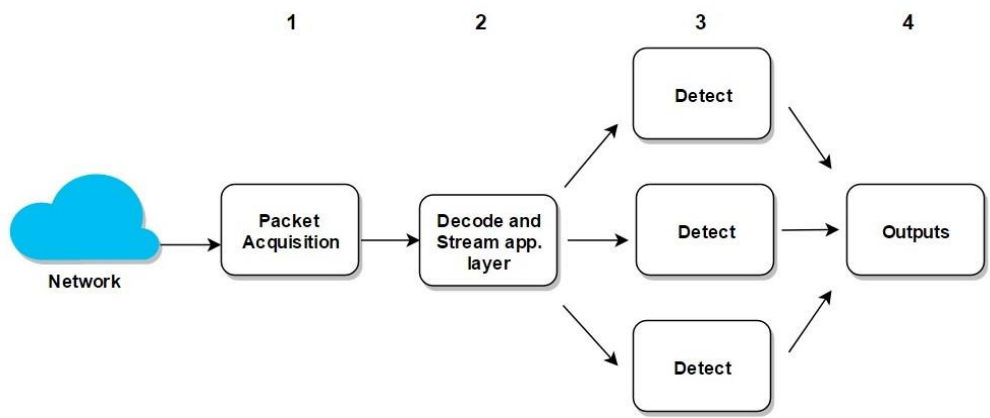

**Figure 2: Suricata multi-threaded architecture**

Snort or Suricata can be installed on a computer host. The performance of both IDSs is directly proportional to the computer host performance in terms of CPU, memory utilisation and the packets drop rate by the network interface card.

The continued increase in network speed and malicious traffic causes significant problems for Snort and Suricata. Both IDSs have to process higher traffic volumes and inspect each incoming network packet to detect malicious traffic. In order for Snort and Suricata to have superior performance, both IDSs must process a vast volume of network traffic that can reach speeds up to 10 Gbps. If both IDSs fail to execute packet inspection at the required rate, they will drop the packets and allow malicious packets to enter the computer network undetected. Therefore, Snort and Suricata should be efficient enough to process network traffic up to 10 Gbps network speed.

Recently, numerous researchers have studied the support of machine learning algorithms for IDSs. Machine learning is a field of computer science that trains the computer to think like humans and take actions where required. In simple processing, a computer processes the information based on statements from primary memory. Machine learning techniques try to copy thinking processes such as logical reasoning, intuition, learning from the past, trial and error and generalisations [6].

Snort and Suricata use rules to detect the known malicious traffic. If malicious traffic matches with the rule set, then they will trigger the alarms. But both IDSs will not take any action against unknown malicious traffic. This is because Snort and Suricata do not use the machine learning techniques and hence cannot stop unknown malicious traffic. Machine learning techniques can help IDSs by accurately detecting the malicious traffic and therefore reduce the false positive alarms by proactively reacting against unknown malicious traffic. There are various machine learning algorithms that can be used for IDSs like Support Vector Machines, Decision Trees, Fuzzy Logic, BayesNet and NaiveBayes.

Snort and Suricata use pre-defined rules to detect malicious network traffic. If malicious traffic patterns match with the rule set then both IDSs trigger alarms, and these can be false positive, false negative or true positive alarms. Snort and Suricata alongside all the other IDSs have a common problem which is triggering false positive alarms [7]. For example, legitimate network traffic consists of DNS or web requests can lead the IDS to trigger a false positive alarm. This is because both IDSs have an identical rule set to match patterns for DNS or Web attacks. Therefore, false positive alarms affect the performance of Snort and Suricata and utilise a high computing resource for classifying the network traffic.

The aim of this paper is to do a performance comparison of Snort and Suricata and to implement machine learning algorithms on it to improve the detection accuracy. The aim led to the following objectives: (1) To critically review Snort and Suricata by measuring the performance and detection accuracy of both IDSs. (2) To evaluate the machine learning algorithms using three different datasets and to improve the selected IDS performance by implementing collective and optimized machine learning techniques through reducing the false positive alarms. The main contribution of this work is the performance comparison of Snort and Suricata at 10 Gbps and the application of the hybrid and optimized machine learning (ML) algorithms to Snort.

## 2. RELATED WORKS ON PERFORMANCE COMPARISON

The idea of a performance comparison between Snort and Suricata is not new. Both perform well, but are not perfect and have limitations as shown in our experiments. Snort has a single-threaded architecture, and Suricata has a multi-threaded architecture which makes both IDS distinct from each other, but the rule set is the common feature of both IDS. Classifying the network traffic and accuracy of the rule set are the key elements of both IDS's performance. Furthermore, computer host performance has a clear impact on the overall IDS performance. A performance comparison study [8] was carried out on Snort and Suricata IDS and the experiments were performed to ascertain computer host resource utilisation performance and detection accuracy. The experiments were performed on two different computer hosts with different CPU, memory and network card specifications. Their results showed that Suricata required a higher processing power to work well generally as compared to Snort. Moreover, the results showed that with higher processing power Suricata could accurately detect malicious traffic on the network and its rule set was effective [8].

Later in 2013, the administrative evaluation of three IDSs by Wang et al. concluded that Snort utilised low computing resources and its rule set accurately classified the legitimate and malicious network traffic. The researchers evaluated the performance of three IDSs in a simulated environment. The environment consisted of physical and virtual computers. The experiment results showed that Snort could have a negative impact on network traffic more than the other two tested IDSs [9].

Bulajoul et al. [10] designed a real network to carry out the experiments that used Snort IDS. This study demonstrated the lack of ability of Snort IDS to process a number of packets at high speed and it dropped packets without accurately analysing them. The study concluded that Snort IDS failed to process high-speed network traffic and the packet drop rate was higher. The researchers introduced a parallel IDS technology to reduce the packet drop rate as a solution. (Waleed, Anne & Mandeep, 2013). The performance of Snort IDS was improved by using dynamic traffic awareness histograms. This study discusses the most effective way to use the order of attack signature rules as well as the order of the rule field. The proposed approach uses the histograms for predicting the next signature rules and rule field orders. The simulation performed showed that the proposed approach significantly improved Snort performance [11].

Saboor et al. evaluated the Snort performance against DDoS. The evaluation methodology consisted of three different hardware configurations. The Snort performance was observed in terms of packet handling and detection accuracy against DDoS on three different hardware configurations. The experiments results showed that Snort packet handling could be improved by using better hardware configurations, but Snort detection capability was not improved by using better hardware [12]. Shahbaz, et al. [13] on the efficiency enhancement of IDS, addresses the problem of dimensionality reduction by proposing an efficient feature selection algorithm that considers the correlation between a subset of features and the behaviour class label.

Alhomoud et al. [14] have tested and analysed the performance of Snort and Suricata. Both were implemented on three different platforms (ESXi virtual server, Linux 2.6 and FreeBSD). The traffic speed of up to 2 Gbps was used in this paper. Albin [15] compared the performance of two open-source intrusion detection systems, Snort and Suricata, by evaluating the speed, memory requirements, and accuracy of the detection engines in a variety of experiments. Zammit [16] implemented an intrusion detection system that uses machine learning techniques to classify traffic generated from honeypot interactions. Huang et al. [17] analysed and implemented the Snort intrusion detection model in a campus network. Victor et al. [18] worked to design an operational model for minimization of false positive IDS alarms, including recurring alarms by the security administrator. White et al. [19] presents a thorough comparison of the performance of Snort and Suricata. They examine the performance of both systems as they scale system resources such as the number of CPU cores, the rule sets used and the workloads processed. There are other works that looks at measuring the intrusion detection capability as in [20], tweaking IDS performance as in [21], parallel design of IDS on many-core processors as in [22], an approach for unifying rule based deep packet inspection as in [23], a Better Snort Intrusion Detection/Prevention System (BSnort) that uses Aho-Corasick automaton as in [24], improving the accuracy of network intrusion detection systems as in [25], boosting throughput of Snort NIDS under Linux as in [26], evaluation studies of three IDS under various attacks and rule sets as in [27] etc.

## 3. DESCRIPTION OF RESEARCH METHODOLOGY

Experiment scenarios were designed to make observations and to take measurements. This study demonstrated rigorous, repeatable, quantitative performance comparisons of both IDSs and evaluated the machine learning algorithms.

The experiments consisted of a test bed which compared Snort and Suricata's detection accuracy in 10 Gbps network speed and with seven different types of malicious traffic. The seven types of malicious traffic were chosen because the rules could be applied consistently to both Snort and Suricata. Moreover, they are the most common types and covered a good number of attacks. The experiments compared the performance of both IDSs by measuring the percentage of CPU, memory utilisation and network packet drop rate. Snort and Suricata's rule set detection accuracy is measured by malicious traffic in a controlled experiment environment and compared the number of false positive, false negative and true positive alarms triggered by each IDS. The normal network traffic for the experiments was produced using three open source network traffic generators, namely Ostinato, NMAP, and NPING [28]. These tools can generate network traffic up to 20 Gbps. The malicious traffic was generated using Kali Linux Metasploit framework [29].

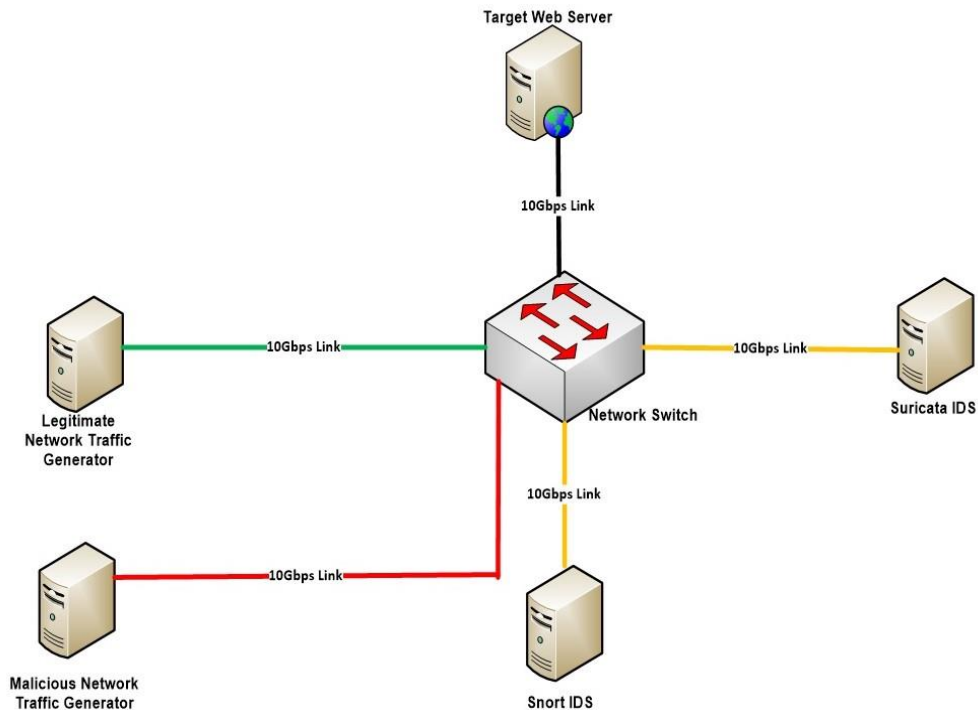

**Figure 3: Experiment Network**

The default rule set of Snort and Suricata were used for the experiments. Also, the legitimate network traffic and malicious traffic were generated and a combined traffic is given as input to Snort and Suricata. Some of the questions we had to ask were as follows. Which IDS has superior performance when processing network traffic up to 10 Gbps? Does the architecture of the IDS have an impact on this? What are the differences in the packet drop rate between Snort and Suricata when the CPU, memory and network utilisation increase? How accurate are Snort and Suricata's rule set, when processing legitimate network traffic alongside malicious traffic?

The experiment network that is shown in Figure 3 was built using Oracle Virtual Box. Five virtual machines (VMs) were created. Depending on the individual experiment requirements, network packets (legitimate and malicious) were produced at varying network speeds with network traffic generator tools. The five VMs were connected via a virtual switch using 10 Gbps Ethernet links. The experiment network consisted of high-performance VMs which were running Snort and Suricata. The latest available version of Snort (v2.9.6.1) and Suricata (v2.6.9) were used for the experiments.

## 4. EXPERIMENT SCENARIOS

The experiment scenarios were planned and setup to compare the performance of Snort and Suricata on identical VMs using identical rule set and under the same test conditions.

### 4.1 Experiment Scenario One

This experiment observed the real-time performance of Snort and Suricata while processing at a 10 Gbps legitimate network speed from a legitimate network traffic generator (Ostinato). The first experiment was carried out to compare Snort to Suricata's performance. To achieve accurate results, the experiment scenario was tested with packets size of 1,470 bytes for TCP, UDP and ICMP. These packets were injected to both IDSs with a network speed of 10 Gbps. The experiment consisted of a logical network diagram as shown in Figure 3. Each IDS was separately installed on identical VMs with default performance parameters and rule set. A number of tools were used to observe and record the measurements of CPU, memory, network utilisation and the packet drop rate. These include Metasploit Framework, Collectl, top, dstat, Snort logs, tcpdump, IPTRAF, nmap etc. and the experiment was conducted for 4 hours. We extended this to two more four-hour blocks, i.e. a total of 12 hours to increase the reliability of our findings. The following packets were injected as the background Snort traffic ranging from 1 Gbps to 10 Gbps network speed as follows. (1) 1,000,000 UDP packets with a rate of 500 packets/sec, each packet size being 1,470 bytes (2) 1,000,000 TCP packets with a rate of 500 packets/sec, each packet size being 1,470 bytes (3) 1,000,000 ICMP packets with a rate of 1000 packets/sec, each packet size being 1,470 bytes.

**Table 1: Experiment network specifications**

| Machine/Software Type | Specification | Tools Used |
|---|---|---|
| CentOS 6.5 | Virtual Machine, 2.5 GHz 4 cores CPU, 4 GB Memory, 10 Gbps Ethernet | Snort 2.9.6.1 IDS; Collectl, top, dstat Snort logs, tcpdump, IPTRAF |
| Malicious Network Traffic Generator | Virtual Machine, 2 cores CPU, 2 GB Memory, 10 Gbps Ethernet | Kali Linux Metasploit Framework |
| CentOS 6.5 | Virtual Machine, 2.5 GHz 4 cores CPU, 4 GB Memory, 10 Gbps Ethernet | Suricata 2.9.6 IDS; Collectl, top, dstat Suricata logs, tcpdump, IPTRAF |
| Legitimate Network Traffic Generator | Virtual Machine, 2 cores CPU, 2 GB Memory, 10 Gbps Ethernet | Ostinato, NMAP, NPING |
| Network Switch | Virtual Switch | |

The collected data from experiment one was recorded and observed. There are options to inject the packets normally or as bursts. The configuration option for burst traffic needs the number of bursts, the packets per burst and bursts per second. We chose normal traffic specifying the number of packets per second and number of packets. Also, multiple instances of packet injection can be made. The data gathered from experiment one showed that Suricata CPU utilisation was higher than that of Snort while processing the same 10 Gbps of network traffic. Table 2 provides the comparison results.

Suricata's average CPU utilisation was 79.3 % for 2.4 GHz on the 4 core CPUs used with different traffic rates, while Snort's CPU sustained an average range utilisation of 67.7 % across the same metrics. Therefore, Snort's individual core CPU utilisation is proved lesser than that of Suricata. The collected performance data showed that Suricata's memory usage was greater than that of Snort as in Table 3. Suricata's average memory usage increased from 3.5 Gbytes when running at 1 Gbps and continued to increase at a variable rate to a maximum of 3.8 Gbytes when processing at 10 Gbps network speed.

**Table 2: CPU utilisation (%) for Snort and Suricata for different traffic rates**

| | First 4 hours | | Second 4 hours | | Third 4 hours | | Average | | Standard deviation | | Confidence Interval (95%) | |
|---|---|---|---|---|---|---|---|---|---|---|---|---|
| Traffic rate | Snort (%) | Suricata (%) | Snort (%) | Suricata (%) | Snort (%) | Suricata (%) | Snort (%) | Suricata (%) | Snort (%) | Suricata (%) | Snort (%) | Suricata (%) |
| 1 Gbps | 50 | 57 | 48 | 52 | 50 | 57 | 49.3 | 55.3 | 0.9 | 2.4 | ±2.3 | ±5.9 |
| 2 Gbps | 52 | 59 | 51 | 59 | 52 | 60 | 51.7 | 59.3 | 0.5 | 0.5 | ±1.2 | ±1.2 |
| 3 Gbps | 54 | 63 | 55 | 61 | 54 | 63 | 54.3 | 62.3 | 0.5 | 0.9 | ±1.2 | ±2.3 |
| 4 Gbps | 57 | 75 | 54 | 71 | 57 | 76 | 56.0 | 74.0 | 1.4 | 2.2 | ±3.5 | ±5.4 |
| 5 Gbps | 65 | 65 | 61 | 68 | 65 | 65 | 63.7 | 66.0 | 1.9 | 1.4 | ±4.7 | ±3.5 |
| 6 Gbps | 63 | 75 | 64 | 70 | 63 | 75 | 63.3 | 73.3 | 0.5 | 2.4 | ±1.2 | ±5.9 |
| 7 Gbps | 70 | 73 | 68 | 70 | 70 | 73 | 69.3 | 72.0 | 0.9 | 1.4 | ±2.3 | ±3.5 |
| 8 Gbps | 63 | 80 | 66 | 77 | 63 | 80 | 64.0 | 79.0 | 1.4 | 1.4 | ±3.5 | ±3.5 |
| 9 Gbps | 70 | 78 | 68 | 74 | 70 | 78 | 69.3 | 76.7 | 0.9 | 1.9 | ±2.3 | ±4.7 |
| 10 Gbps | 69 | 80 | 65 | 78 | 69 | 80 | 67.7 | 79.3 | 1.9 | 0.9 | ±4.7 | ±2.3 |

**Table 3: Memory utilisation (Gbps) for Snort and Suricata for different traffic rates**

| | First 4 hours | | Second 4 hours | | Third 4 hours | | Average | | Standard deviation | | Confidence Interval (95%) | |
|---|---|---|---|---|---|---|---|---|---|---|---|---|
| Traffic rate | Snort | Suricata | Snort | Suricata | Snort | Suricata | Snort | Suricata | Snort | Suricata | Snort | Suricata |
| 1 Gbps | 2.9 | 3.6 | 2.7 | 3.4 | 2.5 | 3.5 | 2.7 | 3.5 | 0.2 | 0.1 | ±0.4 | ±0.2 |
| 2 Gbps | 3.0 | 3.7 | 2.8 | 3.4 | 2.6 | 3.5 | 2.8 | 3.5 | 0.2 | 0.1 | ±0.4 | ±0.3 |
| 3 Gbps | 3.0 | 3.6 | 3.0 | 3.5 | 2.8 | 3.6 | 2.9 | 3.6 | 0.1 | 0.0 | ±0.2 | ±0.1 |
| 4 Gbps | 3.1 | 3.6 | 3.0 | 3.5 | 2.9 | 3.6 | 3.0 | 3.6 | 0.1 | 0.0 | ±0.2 | ±0.1 |
| 5 Gbps | 3.2 | 3.8 | 3.1 | 3.7 | 3.1 | 3.7 | 3.1 | 3.7 | 0.0 | 0.0 | ±0.1 | ±0.1 |
| 6 Gbps | 3.3 | 3.7 | 3.0 | 3.8 | 3.0 | 3.6 | 3.1 | 3.7 | 0.1 | 0.1 | ±0.4 | ±0.2 |
| 7 Gbps | 3.0 | 3.7 | 3.1 | 3.7 | 3.0 | 3.7 | 3.0 | 3.7 | 0.0 | 0.0 | ±0.1 | ±0.0 |
| 8 Gbps | 3.3 | 3.8 | 3.1 | 3.7 | 3.0 | 3.7 | 3.1 | 3.7 | 0.1 | 0.0 | ±0.3 | ±0.1 |
| 9 Gbps | 3.4 | 3.7 | 3.2 | 3.7 | 3.2 | 3.7 | 3.3 | 3.7 | 0.1 | 0.0 | ±0.2 | ±0.0 |
| 10 Gbps | 3.2 | 3.8 | 3.2 | 3.8 | 3.2 | 3.7 | 3.2 | 3.8 | 0.0 | 0.0 | ±0.0 | ±0.1 |

Snort's average memory usage was comparatively less, starting at 2.7 Gbytes at 1 Gbps and continued to perform with reduced memory usage across all network speeds culminating in a memory usage of 3.2 Gbytes while processing at 10 Gbps network speed. Suricata's memory usage is more to do with the multithreaded architecture.

When the network traffic speed started to increase, then the CPU and memory utilisation also started to increase for both IDSs. Snort and Suricata log files statistically showed the process with packets at variable network speeds and the amount of CPU and memory utilisation. Snort's packet processing engine was slower than Suricata's processing engine. The same amount (1,000,000 packets) of UDP, TCP and ICMP packets were injected into both IDSs for a 12-hour time duration (i.e. three 4-hour slots). Suricata showed excellent performance when compared to Snort while processing large amounts of packets. The average processing rate was 82,223 packets/second during the 12-hour period, whereas Snort's average processing rate was 60,866 packets/second during the same timeframe. Table 4 shows the observed packets processing rate between Snort and Suricata.

As shown in Table 5, the network performance remained an issue as Snort dropped more packets at 10 Gbps network speed compared to Suricata. The experiment was done in three blocks of 4 hours as before, and the average value is taken from the three values for UDP, TCP and ICMP for both IDS.

**Table 4: Snort and Suricata packet processing rate (per second)**

| | First 4 hours | | Second 4 hours | | Third 4 hours | | Average processing rate (per sec) | |
|---|---|---|---|---|---|---|---|---|
| Time elapsed (sec) | Suricata | Snort | Suricata | Snort | Suricata | Snort | Suricata | Snort |
| 900 | 50,000 | 30,000 | 55,000 | 33,000 | 52,000 | 31,000 | 52,333.3 | 31,333.3 |
| 1800 | 60,000 | 40,000 | 62,000 | 38,000 | 60,000 | 40,000 | 60,666.7 | 39,333.3 |
| 2700 | 75,000 | 50,000 | 72,000 | 49,000 | 71,000 | 48,000 | 72,666.7 | 49,000.0 |
| 3600 | 90,000 | 65,000 | 85,000 | 60,000 | 88,000 | 62,000 | 87,666.7 | 62,333.3 |
| 7200 | 90,000 | 70,000 | 88,000 | 68,000 | 90,000 | 69,000 | 89,333.3 | 69,000.0 |
| 8100 | 95,000 | 70,000 | 93,000 | 70,000 | 91,000 | 68,000 | 93,000.0 | 69,333.3 |
| 9000 | 90,000 | 72,000 | 91,000 | 71,000 | 92,000 | 70,000 | 91,000.0 | 71,000.0 |
| 10800 | 90,000 | 75,000 | 92,000 | 76,000 | 90,000 | 74,000 | 90,666.7 | 75,000.0 |
| 13500 | 95,000 | 72,000 | 94,000 | 70,000 | 93,000 | 71,000 | 94,000.0 | 71,000.0 |
| 14400 | 90,000 | 70,000 | 92,000 | 72,000 | 91,000 | 72,000 | 91,000.0 | 71,333.3 |

**Table 5: Average packets drop at varying network speed**

| Network Speed | Packet Drop Rates (average) | | | | | |
|---|---|---|---|---|---|---|
| | UDP | | TCP | | ICMP | |
| | Snort | Suricata | Snort | Suricata | Snort | Suricata |
| 1 Gbps | 3.2% | 1.5% | 1.5% | 0% | 7.0% | 3.5% |
| 2 Gbps | 4.1% | 2.0% | 3.1% | 1.0% | 8.2% | 4.1% |
| 3 Gbps | 4.8% | 2.5% | 4.9% | 2.2% | 9.8% | 4.9% |
| 4 Gbps | 6.3% | 2.9% | 7.2% | 3.6% | 10.5% | 5.5% |
| 5 Gbps | 7.5% | 3.7% | 9.5% | 4.1% | 11. %5 | 6.2% |
| 6 Gbps | 9.6% | 5.0% | 12.1% | 5.3% | 13.0% | 6.9% |
| 7 Gbps | 10.8% | 5.9% | 14.8% | 6.2% | 14.2% | 8.0% |
| 8 Gbps | 11.7% | 6.6% | 16.4% | 7.4% | 15. %6 | 8.6% |
| 9 Gbps | 12.8% | 7.2% | 18.2% | 8.0% | 16.1% | 9.1% |
| 10 Gbps | 14.0% | 8.0% | 20.0% | 9.0% | 17.0% | 10.0% |

Snort and Suricata's log files identified that at 10 Gbps network speed, the host operating system (CentOS 6.5 Linux) was accountable for the total number of packets dropped by both IDSs. The analysis was conducted by observing the host operating system event viewer logs and it was discovered that CentOS 6.5 Linux memory buffers were consumed with packets before both IDSs could read the data contained within. When processing at 100 Mbps and 1 Gbps this problem did not occur. This clearly showed that with higher buffer memory, the packet drop could be reduced for 10 Gbps traffic.

## 4.2 Experiment Scenario Two

This experiment was planned to determine how accurately Snort and Suricata rule set inspected the 10 Gbps network traffic to correctly classify the legitimate and malicious traffic. The metrics listed below were used to measure the detection accuracy of both IDSs. We are not considering True Negative Rate as it is not relevant to us.

**False Positive Rate (FPR)**

This is the likelihood that the IDS will trigger an alarm when there is no intrusion. This was measured by FPR percentage. It triggers an excessive rate of false positive alarms when they do not accurately differentiate between legitimate and malicious network traffic [30].

**False Negative Rate (FNR)**

This is the likelihood that the IDS does not trigger an alarm when there is an intrusion. This was measured by FNR percentage. It does not trigger false negative alarms and allow the network traffic to pass through when the IDS do not have the matching rule set for network traffic [30].

**True Positive Rate (TPR)**

This is the likelihood that IDS trigger an alarm when an intrusion is detected. This was measured by TPR percentage. It triggers a true positive alarm when they accurately detect malicious traffic [30].

The experiment tested the accuracy of Snort and Suricata with two test conditions.

1. Measuring false positive, false negative and true positive rates of both IDSs with legitimate traffic.
2. Measuring false positive, false negative and true positive rates of both IDSs with combined legitimate and malicious traffic at a fixed 10 Gbps network speed.

A default rule set and a default performance configuration of both IDSs were used. Snort and Suricata have identical rule sets to identify malicious network traffic. The seven different types of malicious network traffic were injected and common rule set was observed as per the default rule set. The number of common rule set for both IDSs is shown in Table 6.

**Table 6: Number of common rule set**

| No. | Rules and Malicious Traffic Type | Number of Rules |
|-----|----------------------------------|-----------------|
| 1 | SSH | 13 |
| 2 | DoS/DDoS | 69 |
| 3 | FTP | 75 |
| 4 | HTTP | 110 |
| 5 | ICMP | 125 |
| 6 | ARP | 21 |
| 7 | SCAN | 30 |

### 4.2.1 Modeling the attack

This experiment required a target server running HTTP, FTP and SSH services as shown in Figure 3. Using Kali Linux Metasploit framework the seven types of malicious traffic were generated as shown in Table 6, along with the legitimate traffic. They were all injected to both IDSs to simulate the attacks. Finally, each IDS will inspect the legitimate and malicious traffic and will trigger alarms when the input traffic matches the rule set. The number of alarms (false positive, false negative and true positive) will show how accurately Snort and Suricata classifies the network traffic. The Metasploit framework is used to generate malicious traffic with different exploits and payloads. For example, before executing the payload successfully, an exploit should be identified in the victim computer like open ports through a scan. The basic steps for exploiting a system using the Metasploit Framework which provides different exploits and payloads for different operating systems like Windows, Linux or Mac OS are as follows:

1. Selecting and configuring an exploit which will take advantage of the bugs in an operating system. It may be worthwhile to check whether the target is open to this attack.
2. Selecting and setting up a payload which will be the program that will get executed on the target.
3. Selecting an encoding technique to avoid detection by the IDS.
4. Running the exploit.

We can better understand the need for IDS when we understand the way the network attack is done. To run an exploit the information on the attack target system like the operating system and what services it runs, is needed. This information can be collected with port scanning and fingerprinting tools such as Nmap, or other similar tools. The Metasploit framework is modular and we can mix and match different payloads and exploits to achieve the needed results. The examples of Snort and Suricata IDS rules are as follows and they use a similar syntax in this case.

- A simple Snort rule is: alert icmp any any → any any (msg:"ICMP Packet"; sid:477; rev:3;). This rule does alerting when there is an ICMP packet (ping traffic).
- A simple Suricata rule is: alert http any any → any any (msg:"Do not read gossip during work"; content:"Scarlett"; nocase; classtype:policy-violation; sid:1; rev:1;). This rule will cause an alert when there is browsing or http traffic.

The Metasploit architecture utilises distinct libraries, and the most important one is the Ruby Extension (Rex) library as in figure 4.

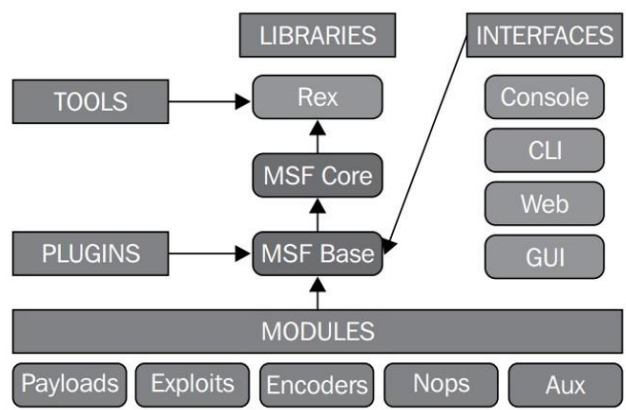

**Figure 4: Metasploit framework architecture [29]**

These libraries are an accumulation of tasks, functions and operations used by the framework. Rex has no dependencies and has the exploitation utility class along with others. The MSF core library extends Rex and allows to communicate with exploit modules. The core library extends the base library which is connected to a different user interface that helps to use command console and the web interface. The later version of the Metasploit framework has around 1412 exploits, 802 auxiliaries, 361 payloads, 327 encoders and 8 nops as shown in the bottom part of figure 4. It is through an exploit that the attacker takes advantage of a flaw in a system, service or application. A payload is the segment of code that can be executed on the exploited system. The auxiliary gives additional options for fuzzing, scanning, recon, DoS attack etc. The encoders are used to obfuscate modules to masquerade and avoid detection by security systems like antivirus or firewall. Nop is No Operation prevents the payload from crashing while executing jump statements in the shell code [29].

A typical network attack pattern can be as in Figure 5. Through a network scan, an attacker can get enough information to execute an exploit remotely, use some stolen information through social engineering or to do a client-side attack.

This could help to escalate the access privileges of the attacker which could lead to stealing of data or exploration of the system. After that, there is also a possibility of pivoting which allows the attacker to move around to other services through exploiting the initial vulnerability. An intrusion detection can alert the administrator when the attacker goes into the second level after the network scan. Some of the performance measurement parameters of the IDS accuracy are as follows.

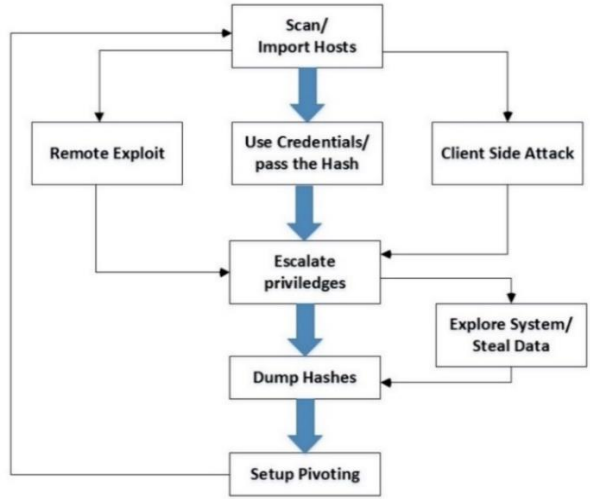

**Figure 5: Network attack pattern [31]**

True Positive (TP): True is identified as true, which is correct identification
True Negative (TN): False is identified as false, which is correct identification
False Positive (FP): True is identified as false, which is wrong identification
False Negative (FN): False is identified as true, which is wrong identification

P = TP + FN (number of correct identification)
N = FP + TN (number of wrong identification)

True Positive Rate (TPR) = TP / P = TP / (TP + FN)                                        (1)
False Positive Rate (FPR) = FP / N = FP / (FP + TN)                                        (2)
False Negative Rate (FNR) = FN / P = FN / (FN + TP)                                        (3)

Experiment two analysed the detection accuracy of Snort and Suricata while processing the legitimate and malicious network traffic. Both the IDSs were kept at the default setting. The first accuracy test was performed using the legitimate network traffic generator which injected UDP, TCP and ICMP packets to both IDSs. The time duration for the first test was 12 hours, i.e. three 4-hour blocks. The log files of Snort and Suricata were analysed, and the results are shown in Table 7. The Suricata's false positive rate (FPR) was higher when processing UDP, TCP and ICMP packets than Snort's FPR. However, Snort did not trigger true positive rate (TPR 0%) and false negative rate (FNR 0%) alarms. In contrast, Suricata triggered a 39% FNR and 3% TPR. Therefore, Snort triggered less false positive alarms. While false negative alarms were observed in both IDSs, Snort's detection accuracy was found to be superior to Suricata in this scenario.

**Table 7: Normal traffic accuracy measurements**

| Normal Traffic | Snort | | | Suricata | | |
|---|---|---|---|---|---|---|
| | FPR | FNR | TPR | FPR | FNR | TPR |
| UDP | 11% | 0% | 0% | 23% | 3% | 0% |
| TCP | 10% | 0% | 0% | 32% | 9% | 0% |
| ICMP | 3% | 0% | 0% | 39% | 27% | 3% |

Both IDSs true positive alarm accuracy was at an acceptable rate and both IDSs were configured with default rule set. The second test was carried out with combined legitimate and malicious traffic at 10 Gbps network speed for about 12 hours, i.e. three 4-hour blocks. Tables 8 and 9 show these results.

**Table 8: Malicious traffic accuracy (%) measurements at 10 Gbps during 12 hours**

| Malicious Traffic | First 4 hours | | | | Second 4 hours | | | | Third 4 hours | | | |
|---|---|---|---|---|---|---|---|---|---|---|---|---|
| | Snort FPR | Snort FNR | Suricata FPR | Suricata FNR | Snort FPR | Snort FNR | Suricata FPR | Suricata FNR | Snort FPR | Snort FNR | Suricata FPR | Suricata FNR |
| SSH | 8.0 | 0.0 | 7.0 | 0.0 | 9.0 | 0.0 | 6.0 | 0.0 | 11.0 | 1.0 | 10.0 | 0.0 |
| DoS/DDoS | 3.0 | 2.0 | 10.0 | 0.0 | 3.5 | 1.0 | 12.0 | 0.0 | 4.0 | 2.0 | 8.0 | 1.0 |
| FTP | 12.0 | 0.0 | 11.0 | 0.0 | 10.0 | 1.0 | 10.0 | 0.0 | 11.0 | 1.0 | 11.0 | 0.0 |
| HTTP | 5.0 | 0.0 | 8.0 | 0.0 | 6.0 | 1.0 | 9.0 | 1.0 | 4.0 | 1.0 | 7.0 | 0.0 |
| ICMP | 20.0 | 0.0 | 22.0 | 0.0 | 17.5 | 1.0 | 22.0 | 0.0 | 17.0 | 0.0 | 20.0 | 0.0 |
| ARP | 7.0 | 0.0 | 10.0 | 12.0 | 8.0 | 1.0 | 11.0 | 15.0 | 5.0 | 1.0 | 10.0 | 12.0 |
| Scan | 1.0 | 4.0 | 5.0 | 3.0 | 1.5 | 2.0 | 7.0 | 2.0 | 2.0 | 1.0 | 7.0 | 4.0 |
| Total | 56.0 | 6.0 | 73.0 | 15.0 | 55.5 | 7.0 | 77.0 | 18.0 | 54.0 | 7.0 | 73.0 | 17.0 |

**Table 9: Malicious traffic accuracy (%) measurements at 10 Gbps showing average, standard deviation and CI**

| Malicious Traffic | Average | | | | Standard deviation | | | | Confidence Interval (95%) | | | |
|---|---|---|---|---|---|---|---|---|---|---|---|---|
| | Snort FPR | Snort FNR | Suricata FPR | Suricata FNR | Snort FPR | Snort FNR | Suricata FPR | Suricata FNR | Snort FPR | Snort FNR | Suricata FPR | Suricata FNR |
| SSH | 9.3 | 0.3 | 7.7 | 0.0 | 1.2 | 0.5 | 1.7 | 0.0 | ±3.1 | ±1.2 | ±4.2 | ±0.0 |
| DoS/DDoS | 3.5 | 1.7 | 10.0 | 0.3 | 0.4 | 0.5 | 1.6 | 0.5 | ±1.0 | ±1.2 | ±4.1 | ±1.2 |
| FTP | 11.0 | 0.7 | 10.7 | 0.0 | 0.8 | 0.5 | 0.5 | 0.0 | ±2.0 | ±1.2 | ±1.2 | ±0.0 |
| HTTP | 5.0 | 0.7 | 8.0 | 0.3 | 0.8 | 0.5 | 0.8 | 0.5 | ±2.0 | ±1.2 | ±2.0 | ±1.2 |
| ICMP | 18.2 | 0.3 | 21.3 | 0.0 | 1.3 | 0.5 | 0.9 | 0.0 | ±3.3 | ±1.2 | ±2.3 | ±0.0 |
| ARP | 6.7 | 0.7 | 10.3 | 13.0 | 1.2 | 0.5 | 0.5 | 1.4 | ±3.1 | ±1.2 | ±1.2 | ±3.5 |
| Scan | 1.5 | 2.3 | 6.3 | 3.0 | 0.4 | 1.2 | 0.9 | 0.8 | ±1.0 | ±3.1 | ±2.3 | ±2.0 |
| Total | 55.2 | 6.7 | 74.3 | 16.7 | 6.3 | 4.1 | 7.0 | 3.2 | ±15.5 | ±10.1 | ±17.3 | ±7.9 |

Seven different types of malicious traffic were injected into Snort and Suricata. The average FPR for Snort is 55.2% and for Suricata is 74.3%. Snort on average triggered 6.7% FNR and Suricata triggered 16.7%. Both IDSs did not detect all the malicious traffic, for the default configuration of rules.

With both IDSs configured at the default setting, Suricata triggered a high rate of false negative alarms. When Suricata's log file was analysed, it was observed that Suricata's rule set did not process the data link layer frames and therefore it was unable to detect these malicious frames, and the majority of false negative alarms were triggered when Suricata was processing the ARP malicious frames. Comparatively, Snort's accuracy was far superior as it triggered fewer false negative alarms.

Snort accurately detected six of the seven types of malicious traffic. The true positive rate, TPR of 99% was recorded for each of the six types of malicious traffic detected. Snort only failed to detect TPR on Scan malicious traffic (with TPR 0%). Suricata only detected the FTP, SSH, HTTP, DoS/DDoS and ICMP malicious traffic and the TPR was 99%. However, Suricata did not accurately detect the ARP and Scan malicious traffic (with TPR 0%). The TPR information is not shown in Tables 8 and 9.

Some of the observations are as follows. Snort is a more accurate IDS compared to Suricata as per our experiments. Being single threaded it does not utilise high computational resources as expected. At a higher speed of 10 Gbps Snort drops more packets than Suricata, but at lower speeds, packet drop rate is reduced. The packet processing rate of Suricata is found to be better than Snort. Snort is found to have lesser CPU and memory utilisation compared to Suricata for different traffic speeds. A problem with Snort is its false positive alarm rate, though better than Suricata. It is crucial that this is solved because the analysis of the false positive alarms consumes the administrator time and computational resources. Our effort now is to reduce this FPR in Snort through machine learning techniques.

## 5. SNORT IDS USING MACHINE LEARNING TECHNIQUES

We will consider Snort to implement machine learning [32]. Snort is dependent on the rule set to detect malicious traffic that gives an accurate description of the known malicious traffic. To reduce FPR, many researchers have used machine learning algorithms to classify normal and malicious traffic. The following machine learning algorithms have been extensively studied like Support Vector Machine, Decision Trees, Fuzzy Logic, BayesNet and NaiveBayes.

Also, various other machine learning algorithms such as Genetic Algorithms, Neural Networks and Sequential Minimal Optimization Algorithm were used to limit false positive alarms. We were interested to select the best-performing ones to be used in conjunction with the Snort to improve performance by reducing the false positive alarms.

## 6. EVALUATION OF MACHINE LEARNING ALGORITHMS (MLAs)

Before using machine learning techniques in conjunction with Snort, it is important to know about high performing machine learning algorithms. Three publicly available datasets were used to conduct performance experiments on MLAs. Initially, the experiment evaluation environment was built which consisted of evaluation setup and data preprocessing using data mining software [33]. Secondly, a literature search was conducted, and a comprehensive set of five high-performing machine learning algorithms (MLAs) were selected based on previous research findings as follows: Support Vector Machine (SVM), Decision Trees (DT), Fuzzy Logic, BayesNet and NaiveBayes. False positive, false negative and true positive alarm rates were the metrics used for the comparison of detection accuracy of selected MLAs.

### 6.1 Related Works on using MLA with IDS

A number of researchers have studied and used machine learning techniques to solve IDS detection problems [34-38]. The research focused on detecting malicious traffic with machine learning algorithms when patterns are not detected by IDS rule set.

Buczak and Guven [39] did a survey of data mining and machine learning methods for cyber security intrusion detection, where they have outlined methods by different authors. Tang, Jiangand and Zhao [40] combined the Information Gain (IG), Support Vector Machine and K-means machine learning algorithms to detect malicious traffic when patterns do not exist in IDS rule set. The analysis was performed with 10% KDD Cup 1999 dataset [41]. The TASVM accurately detected unknown malicious traffic and improved the performance of IDS [40]. Shanmuham and Idris [42] used the Fuzzy Logic machine learning algorithm to accurately detect the misuse and anomaly attacks. The Fuzzy Logic was analysed with KDD 1999 dataset [41]. As a result of the research, the IIDS Framework was proposed. The IIDS Framework improved the Apriori Algorithm that yielded faster rule generation and detection rates for malicious attacks [42].

Firdausi et al. [43] used the machine learning techniques to analyse the behaviour based malware detection. The malware behaviours were analysed with five machine learning algorithms k-Nearest Neigbour (kNN), NaiveBayes, Decision Tree, Support Vector Machine (SVM) and Multilayer Perception Neural Network (MLP). The analysis of experimental results showed that Decision Tree performs well with 95.9% a false positive rate of 2.4%, a precision of 97.3% and an accuracy of 96.8%. It was concluded that with the use of machine learning techniques malware behaviours could be detected accurately [43]. Farid and Rahman [44] proposed an adaptive Bayesian algorithm to accurately classify the different types of attacks. The experiments were performed with benchmark KDD99 dataset. The experiments results showed that adaptive Bayesian has a maximum detection rate (DR) and minimum false positive rate (FPR) as compare to Naïve Bayesian, Neural Network and Support Vector Machine. In addition, adaptive Bayesian Algorithm took 52.8 ms training time and 13.2 ms testing time whereas Naïve Bayesian took 106.7 ms training time and 26.4 ms testing time [44, 45].

To reduce the false positive alarms in Snort, Neuro-Fuzzy techniques were used. The study analysed the NEFCLASS and Jrip with Snort alerts. These alerts were taken as input and learning from training to reduce the false positive and negative alerts. The proposed solution was based on artificial intelligence techniques which reduced the false positive and negative alerts. In the experiments, DARPA 1999 data set [46] and KDD 1999 dataset [41] were used for training and testing purpose [45]. Kolias et al. [47] have done a paper where they thoroughly evaluated the most popular attacks on 802.11 and analyzed their signatures. They also offer a publicly available dataset containing a rich blend of normal and attack traffic against 802.11 networks.

Zhang et al. [48] proposed a distributed intrusion detection system for smart grids (SGDIDS) by developing and deploying an intelligent module, the analyzing module (AM), in multiple layers of the smart grid where they used the support vector machine (SVM) and artificial immune system (AIS) to detect and classify malicious data and possible cyber attacks. Sinclair et al. [49] built an application that enhances domain knowledge with machine learning techniques to create rules for an intrusion detection expert system that uses genetic algorithms and decision trees to automatically generate rules for classifying network connections. Endler et al. [50] used the Basic Security Module auditing tool of Solaris. The anomaly detector consisted of the statistical likelihood analysis of system calls, while the misuse detector was built with a neural network trained on groupings of system calls that shows the benefits of combining both aspects of detection in future intrusion detection techniques. Chen et al. [51] propose to decompose the Snort signature patterns, and these smaller primary pattern sets can be stored along with their associations to allow dynamic signature pattern reconstruction. Fries [52] did the intrusion detection using fuzzy clustering of TCP packet attributes, and Mehmood and Rais [53] used SVM along with ACO feature subset. Other related works that we looked at are as follows. Data mining and machine learning which deals with reducing false positives in intrusion detection as in [54], a software implementation of a genetic algorithm based approach to network intrusion detection as in [55], hybrid of rough set theory and artificial immune recognition system as a solution to decrease false alarm rate in IDS as in [56], practical real-time intrusion detection using machine learning approaches as in [57], an analysis of supervised tree based classifiers for IDS in [58], misuse-based intrusion detection using Bayesian networks as in [59] etc. Apart from intrusion detection, there are many other application areas where machine learning is applied as in Valipour et al. [60], Valipour [61], Valipour et al. [62], Rezaei et al. [63], and Zhao et al. [64].

## 6.2  Evaluation Setup Machine Type

The open source machine learning software Weka was used in our effort to find the best AI classifier or machine learning algorithm. Weka is a collection of machine learning algorithms for data mining tasks. The algorithms can be applied to datasets. Weka is capable of data pre-processing, classification, regression clustering etc. In addition, it also helps in developing new machine learning algorithms. (Weka, 2013). This empirical study was only relevant to the study of detection accuracy of machine learning algorithms. Three publicly available datasets (NSA Snort IDS Alert Logs, DARPA IDS Dataset and NSL-KDD IDS Dataset) were selected and used for the experiments [65]. The details of these datasets are listed in Table 10.

## 6.3  Machine Learning Algorithms

We used the following five algorithms for testing detection accuracy on different IDS datasets, as they performed the best.

**Support Vector Machines (SVM):** SVM is a group of supervised learning methods which have learning algorithms that analyse data and identify the patterns for data classification or regression analysis. The SVM algorithm is based on probabilistic statistical learning theory [67]. It is based on the concept of decision planes that separates between a set of objects having different class memberships that define decision boundaries.

**Decision Tree (DT):** Decision Tree is a non-parametric supervised learning method and is used for classification and regression. DTs build a tree in which each branch shows a probability between a number of possibilities and each leaf shows a decision. The DT algorithms collect information and use it for the purpose of decision making. In decision trees, each level splits the data according to different attributes and the goal is to achieve perfect classification with a minimal number of decisions.

**Fuzzy Logic:** Fuzzy logic is a mathematical method that uses Boolean algebra for data classification. The fuzzy logic variable holds values between 0 and 1 and according to the fuzzy logic developers, it provides flexibility and accuracy when representing the real-world situation. Fuzzy systems with fuzzy if rules have been used in many scenarios and applications. Fuzzy if-then rule or, simply, fuzzy rule plays a central role in FL systems. The idea behind fuzzy logic is to use fuzzy sets for the membership of a variable, where we can have fuzzy propositional logic and fuzzy predicate logic.

**BayesNet:** BayesNet is a probabilistic graphical model. The graphical model is used to represent knowledge about an unknown domain. The graphs represent the random variable between probabilistic dependencies among the corresponding random variables. These conditional dependencies in the graph are predicted by using statistical methods. It reflects the states of some part of a domain that is being modelled and it shows how those states are related or connected by probabilities.

**NaiveBayes:** NaiveBayes algorithm was created to perform classification tasks. Based on the Bayes' theorem, Naive Bayes methods are a set of supervised learning algorithms with the "naive" assumption of independence between every pair of features. Though simple, Naive Bayes can often outperform some sophisticated classification methods [68].

## 6.4 Analysis and Performance Comparison

We used an equal number of default rule sets for both IDSs. For Snort, these rules files are present under the rules directory in the source code tree. The five high performing machine-learning algorithms that were mentioned before were evaluated with three different datasets as in table 10. We did a 10-fold cross validation in Weka and the process produced 10 equally sized sets. Each set is divided into two groups as follows: 90 labelled data that are used for training and 10 labelled data that are used for testing. It used the selected classifier and trains it from the 90-labelled data and did the classification on the 10-labelled data for testing in round one. It did the same thing for round 2 to 10 and produces 9 more classifications, covering all data. It then averaged the performance of the 10 rounds of classifications produced from 10 equally sized (i.e. 90 training and 10 testing) sets.

To compare the machine learning algorithm's performance, FPR, FNR and TPR of each algorithm was recorded. These metrics can be obtained from Weka and are the most important criteria to evaluate the algorithms. The values of the defined metrics (FPR, FNR and TPR) were inserted into the equations listed below as (4) and (5) to select the superior performing algorithm. The detection rate (DR) and false positive rate (FPR) are the two key factors to evaluate the capability of an IDS. The results of the equations are shown in Table 11. The detection rate and detection accuracy are given as follows.

Detection Rate (DR) = TPR / (TPR + FNR)                     (4)

Detection Accuracy (DA) = (TPR + FNR) / (TPR + FNR + FPR)                     (5)

**Table 10: Details of Data Sets**

| Datasets | Dataset Name | Attack Types |
|---|---|---|
| Dataset 1 | NSA Snort IDS Alert Logs (SIDL, 2014) | • MAC Spoofing<br>• DNS Poisoning<br>• IP Spoofing |
| Dataset 2 | DARPA IDS Dataset [46] | • SSH Attacks<br>• FTP Attacks<br>• Scanning Attacks |
| Dataset 3 | NSL-KDD IDS Dataset [66] | • Denial of Service Attack (DoS)<br>• User to Root Attack (U2R)<br>• Remote to Local Attack (R2L)<br>• Probing Attack |

In the selection process, an algorithm will be rejected if its Detection Rate (DR) is poor despite an excellent False Positive Rate (FPR) or Detection Accuracy (DA). Table 11 shows that with three datasets, the Support Vector Machine (SVM) had superior DR and DA when compared to the other algorithms like Decision Trees, Fuzzy Logic, BayesNet and NaiveBayes.

**Table 11: Output of the machine learning algorithms**

| Dataset 1 | | | |
|---|---|---|---|
| **Machine Learning Algorithms** | **DR** | **FPR** | **DA** |
| Support Vector Machines | 96.8% | 0.7% | 95.6% |
| Decision Trees | 79.2% | 2.9% | 82% |
| Fuzzy Logic | 94.5% | 0.2% | 92.3% |
| BayesNet | 65% | 3.5% | 73% |
| NaiveBayes | 62% | 3% | 70% |
| **Dataset 2** | | | |
| **Machine Learning Algorithms** | **DR** | **FPR** | **DA** |
| Support Vector Machine | 97% | 0.5% | 94.2% |
| Decision Tree | 81.1% | 1.9% | 85% |
| Fuzzy Logic | 92% | 1.6% | 94% |
| BayesNet | 63% | 5.1% | 71.2% |
| NaiveBayes | 65% | 6% | 71% |
| **Dataset 3** | | | |
| **Machine Learning Algorithms** | **DR** | **FPR** | **DA** |
| Support Vector Machine | 97.3% | 3.1% | 95.4% |
| Decision Tree | 78% | 10% | 81.2% |
| Fuzzy Logic | 95% | 4% | 94% |
| BayesNet | 69% | 8% | 74% |
| NaiveBayes | 70% | 7.6% | 79% |

The SVM had a high DR average of 96% and a low FPR average of 3%. The DA average of 95% was recorded for SVM. Fuzzy Logic also had good DR and DA when compared to Decision Tree, BayesNet and NaiveBayes. The Fuzzy Logic over the three datasets had a DR average of 93.8% and a low FPR average of 0.92%. The DA average of 93.4% was recorded. Other algorithms like DTs, BayesNet and NaiveBayes had lower DR and higher FPR when using the same three datasets. The DA was also unsatisfactory as compared to SVM and Fuzzy Logic. Snort in conjunction with machine learning algorithms can be flexible in selecting the machine learning techniques that can accurately classify the legitimate and malicious traffic. It also improves Snort DR and reduces the FPR and the result of which is Snort providing a higher DA.

Based on the results in Table 11, the SVM was selected. The SVM is used to solve a binary classification problem in a supervised environment and it can train with a large number of patterns. With the use of hyperplane, SVM segregates instances from different classes and makes sure that all the instances are outside the margin. This produces a hard margin and can be represented as follows:

$$y_i(w.x_i + b) \geq 1 \, for \, 1 \leq i \leq n, w \in R^d, w \in R \tag{6}$$

where $x_i$ denotes instances, $y_i \in \{-1, 1\}$ are labels of instances, an intercept term is $b$, $w$ is normal vector to the hyperplane, $d$ is the dimension of input vector and $n$ is the number of input data.

In the real world scenario, the hyperplane will not be found as data usually have a few outliers, instances that are significantly different from other instances from the same class. To tackle this problem, the soft margin was introduced and is expressed as follows.

$$y_i(w.x_i + b) \geq 1 - \epsilon_i, \epsilon_i \geq 0, 1 \leq i \leq n \tag{7}$$

where $\epsilon_i$ denotes slack variables, which allow instances to fall off the margin. To find the optimal soft margin we need to do the following:

$$\min \frac{1}{2} \parallel w \parallel^2 + C \sum_{i=1}^n \epsilon_i \tag{8}$$

where $C$ is the soft margin cost function, which controls the classification accuracy.

Since we are not dealing with linearly separable data, we need to replace the dot product with kernel function. The Radial basis function (RBF) which satisfies Mercer's condition is the most commonly chosen and it is defined as follows.

$$K(x_i, x_j) = \exp(-\gamma \parallel x_i - x_j \parallel^2) \tag{9}$$

The parameter $\gamma$ shows the influence of single training example reaches with low values showing far reach and high values showing close reach [75].

## 7. PROPOSED SNORT ADAPTIVE PLUG-IN

As shown in the Table 8 and 9, Snort has successfully detected six types of malicious traffic out of seven, but it triggered high false positive alarms (55.2% average). The proposed Snort adaptive plug-in architecture for Snort is to reduce the false positive alarms using machine learning algorithms. The proposed new architecture of Snort IDS is shown in Figure 6.

The proposed Snort adaptive plug-in operates in parallel with the Snort's rule set. The rationale for integrating the Snort adaptive plug-in parallel with the rule set is because the Snort rule set only detects the known malicious traffic. The unknown or variant malicious traffic can be detected by the plug-in which in turn can reduce the false positive alarms. This reduction in false positive alarms will obviously improve the detection accuracy. The idea is to modify the existing Snort architecture shown in Figure 1 and integrate the additional plug-in which results in a new architecture as shown in Figure 6. The pre-processor will input the network traffic into the plug-in and Snort rule set and both will operate in parallel to more accurately detect the malicious traffic. The adaptive plug-in had four components as follows: (1) **Decode Network Packets:** This component decodes the packet data to obtain detailed information: source and destination IP addresses/ports, MAC addresses, Ethernet frame and packet size. (2) **Classify Network Packets:** This differentiates between legitimate and malicious traffic. (3) **Machine Learning Algorithm:** Adaptive Plug-in used SVM, Fuzzy Logic, Decision Tree, hybrid of SVM and Fuzzy Logic, and optimized SVM with firefly algorithm to process the legitimate and malicious traffic. (4) **False Positive Alarm Reduction:** This component further reduced the false positive alarms to send the true positive alarms to Snort log files.

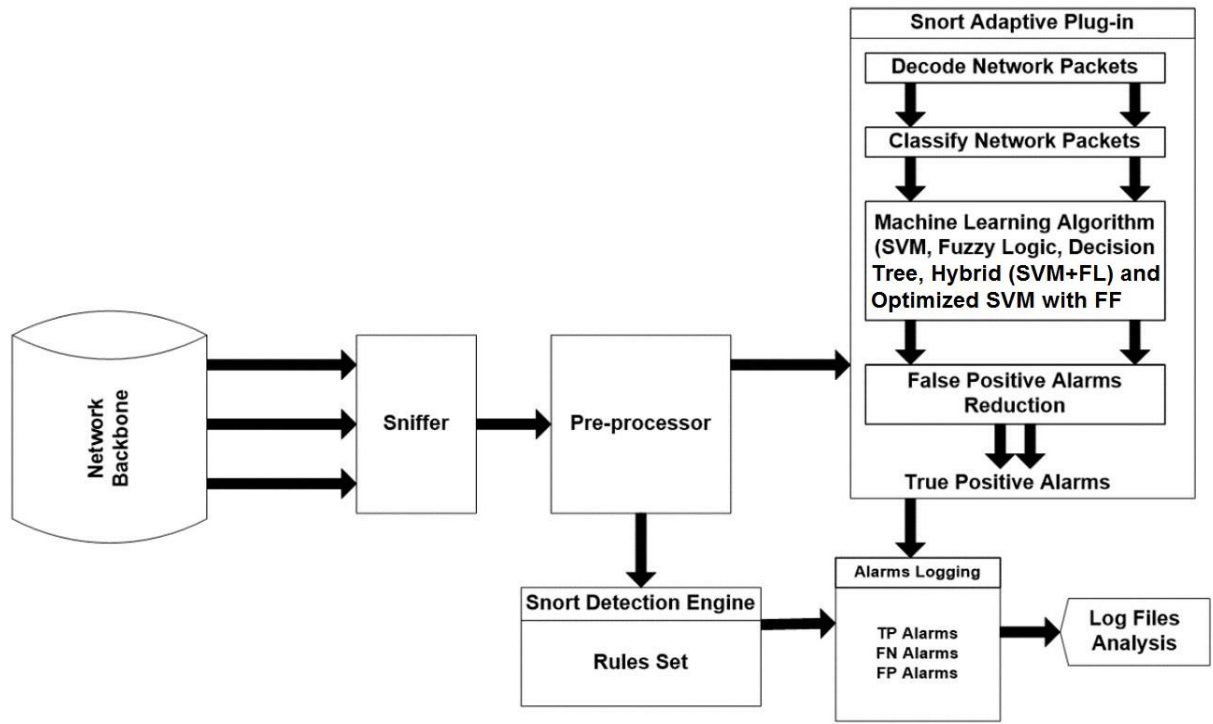

**Figure 6: Proposed Snort IDS Architecture with Snort Adaptive Plug-in**

## 7.1 Snort Adaptive Plug-in Development

The Snort adaptive plug-in for Snort v2.9 intrusion detection system was implemented. The plug-in is a script file written in 'g' programming language (where 'g' is a C programming language in Linux operating system). The plug-in imports the machine learning algorithms, and detects the unknown malicious traffic and reduces false positive alarms. The plug-in operates in parallel with the Snort's detection engine. The pre-processor sends decoded packets to the plug-in. The plug-in uses the machine learning algorithm to classify the legitimate and malicious traffic. Once the network traffic is accurately classified, it will be able to reduce false positive alarms and only send the true positive alarms to Snort's log files.

## 7.2 Evaluation Experiment

The SVM and Fuzzy Logic algorithms showed good performance when evaluated in the Weka engine with three different datasets (as shown in Table 11) and Decision Tree was moderate. However, in order to select the superior MLA: SVM, Fuzzy Logic or Decision Tree, a live background malicious traffic experiment was required for the Snort adaptive plug-in to evaluate the false positive and false negative alarms rates.

## (a) Individual Algorithms

First, the experiments were done for the individual three algorithms using seven types of malicious traffic as shown in Table 7 and 8, which was done three times as 4-hour slots, amounting to 12 hours in total. In SVM, the cost function as in equation (8) was tested with different values and the value of 1.0 was chosen. The gamma values were also tested with different values to change the effect of under-fitting and over-fitting. A gamma value of 0.1 is used. Firstly, the seven types of malicious traffic were injected into the Snort with SVM adaptive plug-in, and later the same traffic was injected into Snort with Fuzzy Logic [69] and Decision Tree plug-in respectively. Table 12 records the FPR and FNR for three cases when they were used with Snort.

**Table 12: Performance of SVM, Fuzzy Logic, and Decision Tree plugins**

| Malicious Traffic | Snort with SVM Plug-in (average values in %) | | Snort with Fuzzy Logic Plug-in (average values in %) | | Snort with Decision Tree Plug-in (average values in %) | |
|---|---|---|---|---|---|---|
| | FPR | FNR | FPR | FNR | FPR | FNR |
| **SSH** | 3.0 | 0.0 | 4.7 | 2.0 | 9.6 | 2.1 |
| **DoS/DDoS** | 1.0 | 1.0 | 7.1 | 0.2 | 8.0 | 1.0 |
| **FTP** | 4.0 | 0.5 | 2.8 | 0.0 | 5.1 | 1.0 |
| **HTTP** | 2.0 | 1.0 | 8.3 | 1.5 | 12.3 | 0.5 |
| **ICMP** | 3.9 | 0.8 | 13.5 | 0.0 | 11.7 | 0.5 |
| **ARP** | 2.0 | 0.0 | 2.0 | 0.0 | 3.2 | 1.0 |
| **Scan** | 1.0 | 0.8 | 1.0 | 0.0 | 2.0 | 1.1 |
| **Total** | **16.9** | **4.1** | **39.4** | **3.7** | **51.9** | **7.2** |

In the initial experiment (refer to Table 8 and 9) Snort was tested without the adaptive plug-in. When seven types of malicious traffic were injected into Snort, it triggered an average value of 55.2% FPR and 6.7% FNR. When Snort was tested with SVM adaptive plug-in as in Table 12, it only triggered 16.9% FPR and 4.1% FNR, compared to Fuzzy logic and DT. This shows that the detection accuracy has really improved.

**(b) Hybrid and Optimized Algorithms**

Second, the experiment was tried with a hybrid version of SVM with Fuzzy Logic [70, 71] and an optimized version of SVM using firefly algorithm [72,73]. Table 13 records the FPR and FNR for the two improved cases when they were used with Snort.

**Table 13: Performance of Hybrid (SVM + Fuzzy logic) and Optimized SVM with Firefly plugins**

| Malicious Traffic | Snort with SVM and Fuzzy Logic Hybrid Plug-in (average values in %) | | Snort with Optimized SVM with Firefly Plug-in (average values in %) | |
|---|---|---|---|---|
| | FPR | FNR | FPR | FNR |
| **SSH** | 2.0 | 0.0 | 1.6 | 0.1 |
| **DoS/DDoS** | 1.0 | 0.5 | 1.0 | 0.2 |
| **FTP** | 3.0 | 0.5 | 2.0 | 0.2 |
| **HTTP** | 2.0 | 1.0 | 1.5 | 0.9 |
| **ICMP** | 2 | 0.7 | 1 | 0.5 |
| **ARP** | 2.0 | 0.0 | 1.0 | 0.0 |
| **Scan** | 1.0 | 0.5 | 0.5 | 0.3 |
| **Total** | **13** | **3.2** | **8.6** | **2.2** |

Under the SVM with the Fuzzy Logic approach, for each given input the basic SVM takes a set of input data and predicts two possible classes of output. That makes it a binary linear classifier that is non-probabilistic in nature. From the output of SVM, the decision-making rule of fuzzy logic is used and the results generated. The average values of the four runs of the experiment are taken as the standard deviation was small as in the previous table 9 without the plugin. With the hybrid version of SVM and Fuzzy logic implemented, the FNR was 13% and FNR was 3.2%. This is a better result than the individual ones in Table 12.

We also optimized SVM classifier with firefly algorithm [74] and it produced the best result compared to the others. SVM parameters (average value) determined by firefly algorithm are as follows: C = 1.57 and γ = 0.58. The firefly algorithm is inspired by the flashing behaviour of fireflies and is a metaheuristic algorithm initially proposed by Xin-She Yang. Since all the fireflies are considered unisex they are not attracted based on the sex but through their brightness. The less bright firefly will move towards the brighter one, with the brightness reversely proportional to the distance.

The brightness of a firefly at a given location $x$ is given as follows.

$$I(x) = \begin{cases} \frac{1}{f(x)} & if\ f(x) > 0 \\ 1 + |f(x)| & otherwise \end{cases} \qquad (10)$$

The attractiveness of a firefly ($\beta$) depends on the distance between the firefly and the one who looks at it, which is proportional to the light intensity of the firefly. It is shown as follows.

$$\beta(r) = \frac{\beta_0}{1 + \gamma r^2} \qquad (11)$$

The position of a firefly $i$ attracted to another brighter firefly $j$ is as follows.

$$x_i^{t+1} = x_i^t + \beta e^{\gamma r^2} ij \left( x_j^t - x_j^t \right) + \alpha_t \epsilon_i^t \qquad (12)$$

where α is randomization parameter, $\epsilon_i^t$ is a vector of random numbers drawn from a Gaussian distribution or uniform distribution at time $t$, and $r_{ij}$ is distance between fireflies $i$ and $j$. The distance between fireflies $i$ and $j$ is calculated using Cartesian distance. The firefly is used to optimize the values of $C$ and $\gamma$ for SVM. Thus, the optimized parameters of SVM are calculated through firefly algorithm and is given as input to the SVM classifier. Each firefly is compared against the other and the best location is selected based on the firefly's brightness. The FPR was 8.6% and FNR was 2.2%, which is the best result that we have achieved.

**Table 13: CPU and Memory utilisation at varying network speed for Snort with adaptive plugin**

| Network Speed | CPU Utilisation | | Memory Utilisation | |
|---|---|---|---|---|
| | Snort without plugin (baseline) | Snort with optimized SVM plugin | Snort without plugin (baseline) | Snort with optimized SVM plugin |
| 100 Mbps (low) | 30% | 35% | 1.5 Gbytes | 1.6 Gbytes |
| 1 Gbps (high) | 45% | 50% | 2.3 Gbytes | 2.6 Gbytes |
| 10 Gbps (very high) | 65% | 73% | 3 Gbytes | 3.7 Gbytes |

As in Table 13 the performance with the optimized SVM plugin is more resource intensive with higher CPU and memory utilizations, which can be tackled with better hardware like more memory and faster mufti-core CPU. We also observed that the speed of the detection with and without plugin are comparable and did not observe any noticeable delay.

## 8. CONCLUSION

This study compared the performance of two open source IDSs, namely Snort and Suricata. They were both proved to be efficient and high performing IDS, though each had its own strengths and weaknesses. The analysis of the experiment results showed that Snort utilised less computational resources to process 10 Gbps network traffic whereas Suricata's utilisation was higher. Suricata processed a higher number of packets per second as compared to Snort, and both IDSs had a high rate of false positives alarms.

Experiment one showed that as a result of Suricata's multi-threaded architecture it required more memory and CPU resources than Snort. It was observed that the 4 cores CPU utilisation of Suricata was 10-15% higher than that of Snort. Suricata used an average 3.8 Gbyte of memory which exceeded Snort's memory utilisation by 600 Mbytes at 10 Gbps. On the analysis of Suricata's log files, it was observed that Suricata required higher computational resources to process network traffic due to the 3 detection stages in built within. The experiment results demonstrated that Suricata had a high ability to process 10 Gbps of network speed with minimum packet drop rate. It further showed that it could process high numbers of packets per second as compared to Snort. Snort is a lightweight single architecture IDS which utilised lower computational resources however it was less efficient in processing 10 Gbps of network speed. Furthermore, the experiment showed Snort had 60,866 packets/second average processing rate while Suricata's processing rate was superior at 82,223 packets/second. It was observed that to process an equal number of packets Snort required greater amount of memory to accommodate the huge volume of network traffic. Snort has added a multi-instance feature to its 2.9 release to address the limitation of single-thread and the version 3.0 stable release will be multithreaded by default. So, the performance can be better for Snort in the future.

Experiment two focused on the importance of detection accuracy of both IDSs. When legitimate and malicious traffic were combined and injected into Snort and Suricata, both IDSs triggered a high rate of false positive alarms. Snort triggered 55.2% FPR compared to Suricata's 74.3% FPR. The default rule set was configured for both IDSs to detect 7 types of malicious traffic; however, the rules used resulted in a high rate of false positive alarms being triggered. Comparatively Snort's FPR detection far exceeded that of Suricata. Both IDSs processed each type of attack differently, and the results showed Snort had a high detection accuracy and triggered only 6.7% FNR whereas Suricata triggered 16.7% FNR while processing the same malicious traffic.

Snort rule set successfully detected six of the seven different types of malicious traffic as compared to Suricata's four. Snort and Suricata's true positive detection rate was 99% in each type of malicious traffic detected. Upon further investigation of Suricata's log files, it was observed that Suricata detected fewer types of malicious traffic due to its incapability of detecting data link layer traffic. Snort has proved itself to be the superior IDS in various aspects however it still triggers a high rate of false positive alarms. To mitigate this problem, machine learning techniques were introduced and an adaptive plug-in was developed to operate in parallel with the Snort rule set. The empirical study conducted sets out to find which machine learning algorithm (MLA) is superior in reducing the false positive alarms. The performance of five different MLAs was compared by using three different datasets in Weka data mining software. The results demonstrated that Support Vector Machine (SVM) was the superior MLA as it reduced the false positive rate to a greater degree than the other MLAs evaluated. SVM showed a high detection rate average of 96% and a low false positive rate average of 3%. A detection accuracy average of 95% was recorded for SVM. The Snort adaptive plug-in initially used SVM to reduce the false positive alarms (to 16.9%). The hybrid approach of SVM with Fuzzy logic produced a better detection accuracy with the FPR as 13% and FNR as 3.2%. The best result was achieved while using optimized SVM with firefly algorithm with the FPR as 8.6% and FNR as 2.2%. We would like to extend this work further for other intrusion detection systems and with other possible hybrid approaches to machine learning and fine-tuning of parameters where possible.

**Author Biographies:**

Syed Ali Razah Shah has finished MSc in the School of Computing, Teesside University, UK and he has research interest in computer networks, network security and AI based applications.

Dr Biju Issac is a senior lecturer in the School of Computing, Teesside University, UK. He has done Bachelor of Engineering in Electronics and Communication Engineering (ECE), Master of Computer Applications (MCA) with honours and PhD in Networking and Mobile Communications, by research. He is a Charted Engineer (CEng), and Senior Member of IEEE.